\documentclass[journal]{IEEEtran}

\usepackage{graphicx}
\usepackage{amsmath,amssymb,amsfonts}
\usepackage{algorithmic}
\usepackage{graphicx}
\usepackage{textcomp}
\usepackage{cite}
\usepackage{flushend, cuted}
\usepackage{lipsum}
\usepackage{stfloats}
\usepackage{cases}
\usepackage{multirow}
\usepackage{booktabs}
\usepackage{balance}
\usepackage{enumitem}
\usepackage{inconsolata}
\usepackage{bbm}
\usepackage{color}
\usepackage{xcolor}
\usepackage{bm}
\usepackage{hyperref}
\usepackage{threeparttable}
\usepackage{makecell}
\usepackage{booktabs} 
\usepackage[ruled,linesnumbered]{algorithm2e}
\usepackage{subcaption}

\ifCLASSINFOpdf

\else

\fi
\begin{document}

\title{
{An Extended State Space Model of Aggregated Electric Vehicles for Flexibility Estimation and Power Control}
\thanks{This work was supported by Natural Sciences and Engineering Research Council (NSERC) Discovery Grant, NSERC RGPIN-2022-03236 and by Fonds de recherche du Québec under Grant FRQ-NT PR-298827.}
}

\author{Yiping Liu,~\IEEEmembership{Student Member,~IEEE}, Xiaozhe Wang,~\IEEEmembership{Senior Member,~IEEE}, Geza Joos, ~\IEEEmembership{Life Fellow,~IEEE
}}

\maketitle

\begin{abstract}
The increasing penetration of electric vehicles (EVs) can provide substantial electricity to the grid, supporting the grids' stability.  
The state space model (SSM) has been proposed 
for power prediction and centralized control of aggregated EVs, offering  low communication requirements and computational complexity. However, the SSM may overlook specific scenarios, leading to significant prediction and control inaccuracies.
This paper proposes an extended state space model (eSSM) for aggregated EVs and develops associated control strategies. By accounting for the limited flexibility of fully charged and discharged EVs, the eSSM more accurately captures the state transition dynamics of EVs in various states of charge (SOC). 
Comprehensive simulations show that the eSSM will provide more accurate predictions of the flexibility and power trajectories of aggregated EVs, and more effectively tracks real-time power references compared to the conventional SSM method. 
\end{abstract}


\IEEEpeerreviewmaketitle
\section{Introduction}
 \IEEEPARstart{A}{s} a clean 
 means of transportation, EVs are increasingly being integrated, contributing to reducing humanity's carbon footprint. 
Despite the challenges posed by their random high power demand, 
EVs, as a mobile energy storage resource, can provide multiple ancillary services to the grid  \cite{gong2023data}. 
Unlike conventional generators, EVs can rapidly adjust their output due to their fast ramping capabilities. 
By controlling the bidirectional flow of power between EVs and the grid, they can meet various ancillary service needs. While a single EV’s regulation capacity may be limited, aggregating multiple EVs through an electric vehicle aggregator (EVA) \cite{kiani2022extended} can deliver substantial regulation power. The "Vehicle-to-Grid" (V2G) technology enables EVs to provide/absorb power to/from the grid for stability support and ancillary service. 
By switching the EVs between different states, 
EVs become active players for grid stability and play important roles in providing energy-based ancillary service, capacity provision, and load shaving.
The previous works \cite{izadkhast2014aggregate}\cite{kazemtarghi2022optimal}\cite{wang2019state} implement 
control on aggregated EVs for frequency regulation service. The work \cite{rafique2021ev} utilizes EVs for peak demand management.  
In \cite{quiros2015control}\cite{wang2018intelligent}, EVs are combined with thermostically controlled loads for demand response. 


The existing EVA models can be classified into distributed and centralized approaches. In distributed methods \cite{kiani2022extended} \cite{gusrialdi2017distributed}\cite{bayram2021pricing}, the control authority is decentralized to individual EVs, enabling each EV 
to optimally schedule its power output and iteratively update its power trajectories to the aggregator with some intelligent devices \cite{yu2023communication}. In contrast, a centralized scheme involves the EVA collecting all available information from each EV and dispatching control signals to determine their states. By making decisions based on global constraints and system-level preferences, centralized control ensures coordinated operation across all EVs, reducing energy losses. However, as the number of EVs increases, the computation and communication complexity required for data collection and transmission also grows significantly \cite{song2018state}. 
To address this issue, references \cite{wang2019state,wang2020electric,kiani2022extended} propose a state space model (SSM) that classifies large EV populations into different state intervals according to their states of charge (SOC).  Control strategy can be developed to control the transition of different intervals to regulate frequency and energy imbalances. As a result, the computation complexity is no longer dependent on 
the population of EVs 
but on the number of state intervals. 
While the SSM successfully reduces the computational demands and achieves high accuracy in  
power trajectory prediction, it 
overlooks the scenarios where EVs are fully charged and discharged. This limitation can cause significant prediction errors in the flexibility of EVs 
and in turn influence the control of aggregated EVs.


In light of the challenge, we propose an extended SSM (eSSM) that incorporates the state transitions of fully charged and discharged EVs. Our simulation results show that the eSSM can provide more accurate  
predictions of the flexibility and power trajectories of aggregated EVs compared to the SSM \cite{wang2019state}. Additionally, the control framework
based on the proposed eSSM can better follow the real-time power control reference, effectively meeting the power requirement for ancillary services such as frequency regulation. 


\color{black}

\section{Modeling Framework}
In this section, we first present the state model of an individual EV, followed by the proposed eSSM model. Compared to the original SSM model \cite{wang2019state}, the eSSM considers the fully charged and discharged EVs by adding two extra states. The formulation and detailed calculation of the eSSM are presented to highlight its advantages.\color{black}

\subsection{State Model of an Individual EV}
Based on the direction of an EV's power flow, 
the connection state of an EV could be classified into: (i) charging state (CS): the EV is withdrawing active power from the grid, (ii) idle states (IS): there is no active power exchange between the EV and system, and (iii) discharging states (DS): the EV is discharging active power to the system with rated discharging power\cite{wang2019state}. Based on the states, the state model of an individual EV$_i$ can be represented as \eqref{eq:cal_S}:
\begin{equation}
\begin{split}
\begin{small}
S_{i}(t+\Delta t) = \begin{cases}
S_{i}(t) + P_{i}(t)\cdot \eta_{i} /Q_{i} \cdot \Delta t, &P_{i}(t) = P_{c,i} \\
S_{i}(t), &P_{i}(t) = 0 \\
S_{i}(t) - P_{i}(t)/\eta_{i}/Q_{i} \cdot \Delta t, &P_{i}(t) = P_{d,i}, 
\end{cases}
\end{small}
\end{split}
\label{eq:cal_S}
\end{equation}
where 
$ \Delta t $ is the time interval; $S_{i}(t)$ is the SOC at the time $ t $; $Q_{i}$ is the battery capacity; $P_{c,i}$ and $P_{d,i}$ are the rated charging and discharging power; $\eta_{c,i}$ and $\eta_{d,i}$ are the charging and discharging efficiencies; $P_{i}(t)$ is the power output.

\begin{figure}[!ht]
\vspace{-0.2cm}
    \centering
    \includegraphics[width=0.6\linewidth]{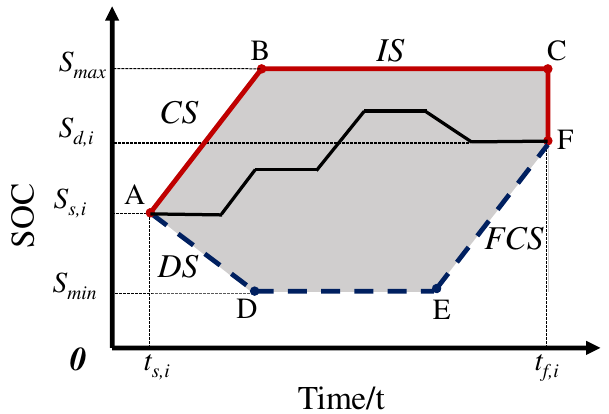}
    \caption{Operation area of an individual EV\cite{wang2019state}}
    \label{fig:operation}
\end{figure}

Fig. \ref{fig:operation} shows the operation area of an individual EV. $t_{s,i}$ and $t_{f,i}$ are the plugging in and plugging out time; $S_{min}$ and $S_{max}$ are the minimum and maximum SOC value. The EV plugs into the grid at point $A$ with initial SOC value $S_{s,i}$ at $t_{s,i}$, and $S_{d,i}$ is the minimum demanded SOC value when the EV plugs out at $t_{f,i}$. 
The upper bound $'A-B-C'$ denotes that the EV enters CS immediately upon plugging in and remains charging until it reaches $S_{max}$. The lower bound $'A-D-E-F'$ indicates that it starts to discharge until it decreases to $S_{min}$. Specifically, to ensure $S_{d,i}$ is attained at $t_{f,i}$, the EV may enter the force charging state (FCS) ($'E-F'$). When connecting to the grid, 
the EV can operate between these two bounds.

According to their properties, the parameters of EVs are categorized into (i) characteristic parameters: $P_{c,i}$, $P_{d,i}$, $Q_{i}$, $\eta_{c,i}$, $\eta_{d,i}$; (ii) traveling parameters: $t_{s,i}$, $t_{f,i}$, $S_{min}$, $S_{max}$, $S_{s,i}$, $S_{d,i}$; and (iii) operation parameters: $S_{i}(t)$, $P_{i}(t)$. 


\subsection{{Extended} State Space Model of Aggregated EVs}

The SSM 
\cite{wang2019state} splits the SOC range $[S_{min}, S_{max}]$ into $N$ intervals as shown in Fig \ref{fig:ControlS}. A connected EV can find its location in the space by its connecting state and SOC value. EVs move in the state intervals with a rated speed under \eqref{eq:cal_S}. 
As discussed in \cite{wang2019state}, the EV may reach the FCS $\left('E-F'\right)$ to ensure the demanded SOC value $S_{d,i}$ is reached at $t_{f,i}$. In this case, it is forced to get charged until plugging out and loses its regulation flexibility. Nevertheless, there are another two special cases that need to be handled separately, which were not considered in \cite{wang2019state}: \color{black}
(i) an EV in CS reaching its maximum SOC ($S_{max}$) transitions to IS; (ii) an EV in DS reaching its minimum SOC ($S_{min}$) transitions to IS. 
The EVs in 
the aforementioned two conditions will lose 
their charging and discharging flexibility, respectively. 

For convenience of expression, let $ N_{s} = N+1$. The real-time distribution of aggregated EVs can be denoted by a state vector $\bm{x}(t)\in \mathbb{R}^{3N_{s}\times 1}$. For 
$i\in\{1,2,..,3N\}$, $x_{i}$  represents the proportion of EVs in the regular state $i$. $x_{3N+1}$ and $x_{3N+2}$ represent the proportion of EVs in IS with $S_{min}$ and $S_{max}$, respectively, while $x_{3N+3}$ represents the proportion of EVs in FCS. The structure of the state intervals is illustrated in Fig. \ref{fig:ControlS}. The blue arrows represent automatic state transitions, while the brown arrows indicate SOC boundary transitions. Control signals are shown by green and red arrows.
\color{black}

Assuming EVs are uniformly distributed among state intervals, the whole movement of EVs can be denoted by a Markov transition matrix $\bm{A}\in \mathbb{R}^{3N_{s}\times{3N_{s}}}$, where $\bm{A}_{m,n}$ indicates the transition probability from state interval $n$ to interval $m$, 
calculated by \eqref{eq:cal_A} and \eqref{eq:cal_P}:
\begin{align}
\vspace{-0.1cm}
    \bm{A}_{m,n} &= \int_{S_{m-1}}^{S_{m}}\int_{S_{n-1}-S_{x}}^{S_{n+1}-S_{x}}P(S_{y}|S_{x}) \label{eq:cal_A}\\
    P(S_{m}|S_{n})&=P(\Delta S) = f(P_{c}, \eta_{c}, Q_{c}) \label{eq:cal_P}
    \vspace{-0.2cm}
\end{align}

where $P(S_{y}|S_{x})$ is defined as the EV's transition probability from SOC value $S_{y}$ to $S_{x}$, $\Delta S$ is the SOC variation every time interval, and $f(P_{c}, \eta_{c}, Q)$ is the joint probability density function of characteristic parameters $P_{c}$, $\eta_{c}$ and $Q_{c}$.

\begin{figure}
    \centering
    \vspace{-0.3cm}
    \includegraphics[width=0.65\columnwidth]{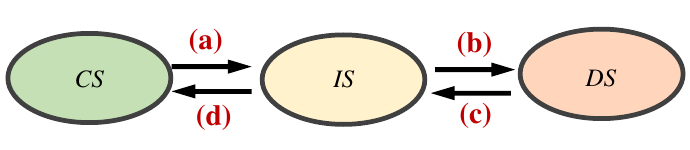}
    \caption{Responding modes of EVs}
    \label{fig:responding_modes}
    \vspace{-0.5cm}
\end{figure}
Assume an EV can only switch between two sequential intervals every time interval. There are four responding modes between connecting states: (a) 'CS' $\rightarrow$ 'IS'; (b) 'IS $\rightarrow$ DS'; (c) 'DS $\rightarrow$ IS'; and (d) 'IS $\rightarrow$ CS'. 'CS $\rightarrow$ DS' is the combination of 'CS $\rightarrow$ IS' and 'IS $\rightarrow$ DS'. The state space model can be structured as follows: 
\vspace{-0.1cm}
\begin{equation}
\begin{small}
    \begin{cases}
\bm{x}(k+1) = \bm{Ax}(k) + \bm{Bu}(k) 
{+\bm{w}(k)}\\
\bm{y}(k) = \bm{Cx}(k)  + \bm{v}(k)
\end{cases}
\label{eq:Extended_SSM}
\end{small}
\vspace{-0.1cm}
\end{equation}
where 
{$\bm{x}(k)$ is the state vector. $\bm{u}(k)\in\mathbb{R}^{4N_{s}}$ is the input vector defined by \eqref{eq:def_u}:
\begin{equation}
\begin{small}
    \bm{u}(k) = 
\begin{bmatrix}
    \bm{u}^{a} & \bm{u}^{b} & \bm{u}^{c} & \bm{u}^{d} & u^{d}_{N_{s}} & u^{b}_{N_{s}} 
\end{bmatrix}^{T}(k)
\end{small}
\label{eq:def_u}
\end{equation}
$\bm{u}^{a}(k)$, $\bm{u}^{b}(k)$, $\bm{u}^{c}(k)$ and $\bm{u}^{d}(k)$  are control inputs, 
indicating the responding modes between connecting state (a)-(d) (see Fig. \ref{fig:responding_modes}). $\bm{u}_{N_{s}}^{b}$ is the input element designed for the state transition between the extended state $S_{max}$ in IS ($\bm{x}_{3N+2}$) and $S_{max}$ in DS ($\bm{x}_{3N}$). $\bm{u}_{N_{s}}^{d}$ is the input vector indicates the state transition between the extended state $S_{min}$ in IS ($\bm{x}_{3N+1}$) and $S_{max}$ in CS ($\bm{x}_{1}$). The domain of each input element is limited by the corresponding value of states. I.e., if $\bm{u}^{a}_{k}(t) > 0$, $\bm{u}^{a}_{k}(t) \leq \bm{x}_{k}(t)$. 
Since the EVs in FCS have 
neither charging nor discharging \color{black} regulation flexibility, there is no input element designed for those EVs. 
\vspace{-0.1cm}
\begin{equation}
\begin{small}
            \bm{B} = 
\begin{bmatrix}
    -\bm{I}_{N\times N} & \bm{O}_{N\times N} & \bm{O}_{N\times N} & \bm{I}_{N\times N}  & \multirow{3}{*}{\makecell{1 \\ \vdots \\ 0}}& \multirow{3}{*}{\makecell{0 \\ \vdots \\ 1}}\\
    \bm{I}_{N\times N} & -\bm{I}_{N\times N} & \bm{I}_{N\times N} & -\bm{I}_{N\times N} &  & \\
    \bm{O}_{N\times N} & \bm{I}_{N\times N} & -\bm{I}_{N\times N} & \bm{O}_{N\times N} &  & \\
    \\
     \bm{O}_{1\times N} & \bm{O}_{1\times N} & \bm{O}_{1\times N} &  \bm{O}_{1 \times N} & -1 & 0 \\
    \bm{O}_{1 \times N} & \bm{O}_{1\times N} & \bm{O}_{1 \times N} & \bm{O}_{1 \times N} & 0 & -1\\
    \bm{O}_{1\times N} & \bm{O}_{1\times N} & \bm{O}_{1\times N} & \bm{O}_{1\times N} & 0 & 0
\end{bmatrix}
\label{eq:def_B}
\end{small}
\end{equation}

where $\bm{I}_{N\times N}$ is a $N\times N$ identity matrix; $\bm{O}_{N\times N}$ is a zero matrix. The element $\bm{B}_{m,n}$ represents the control action $\bm{u}_{n}$ functioning on the state variable $\bm{x}_{m}$, resulting in either a reduction or an increase of the EV population in state $m$. 
For example, if $\bm{B}_{m,n} = 1$, the corresponding state change is $\bm{x}_{m}(k+1)=\sum\limits_{j=0}^{j=3N_{s}}\bm{A}_{m,j}\bm{x}_{j}(k)+\bm{u}_{n}(k)$, increasing the EV population in state $m$.  

$\bm{C}\in\mathbb{R}^{3\times 3N_{s}}$ is a constant matrix given in \eqref{eq:def_C}: 
\begin{equation}
\begin{small}
\bm{C} = N_{EV}
\begin{bmatrix}
    -P_{ac}\bm{1}_{1\times N} & P_{ad}\bm{1}_{1\times N} & -P_{ac}\bm{1}_{1\times N} \\
    \bm{O}_{1\times N} & P_{ad}\bm{1}_{1\times N} & -P_{ac}\bm{1}_{1\times N} \\
    P_{ad}\bm{1}_{1\times N} & P_{ad}\bm{1}_{1\times N} & -P_{ac}\bm{1}_{1\times N} \\
    0 & 0 & -P_{ac} \\
    0 & P_{ad} & 0 \\
    -P_{ac} &  -P_{ac} & -P_{ac}
\end{bmatrix}^{T}
\label{eq:def_C}
\end{small}
\end{equation}
\color{black}
where 
$N_{EV}$ is the total number of connected EVs; \color{black}$P_{ac}$ is the average rated charging power of EVs in CS; $P_{ad}$ is the average rated discharging power in DS; $\bm{1}$ represents a matrix of all ones with size given by the relative subscript.

$\bm{y}(k)$ is the output vector that indicates the flexibility and power trajectory of aggregated EVs as \eqref{eq:output_vector}:
\vspace{-0.1cm}
\begin{equation}
    \begin{small}
        \bm{y}(k) = 
    \begin{bmatrix}
        P_{EV}(k)&
        P_{u}(k) &
        P_{l}(k)
    \end{bmatrix}^{T}
    \end{small}
    \label{eq:output_vector}
    \vspace{-0.1cm}
\end{equation}$P_{EV}(k)$ is the total power output from the aggregated EVs to the grid in the V2G mode. 
$P_{u}(k)/P_{l}(k)$ is the upper/lower bound of power that the aggregated EVs can provide to the grid, which represents the flexibility of aggregated EVs. 
$\bm{w}(k)$ is the noise vector determined by the plugging-in and plugging-out behavior of EVs, calculated by \eqref{eq:noise_vector} \cite{wang2019state}:
\begin{equation}
    \begin{small}
        \bm{w}(k) = \frac{N_{in}\bm{x}_{in} - N_{out}\bm{x}_{out}}{N_{EV} + N_{in} - N_{out}}
    \end{small}
    \label{eq:noise_vector}
\end{equation}
where $N_{in}$ and $N_{out}$ are the number of EVs plugging in and out during the time interval, respectively; $\bm{x}_{in}$ and $\bm{x}_{out}$ are the state distributions of plugging-in and plugging-out EVs, respectively.
Lastly, $\bm{v}(k)$ 
represents a combined measurement noise and modeling error, which is assumed to be an independent Gaussian random vector, i.e. $\bm{v}(k) \sim \mathcal{N}(\bm{0}, \bm{\Sigma}_{\bm{v}})$.
\color{black}



\begin{figure}[t]
\centering
\includegraphics[width=1\columnwidth]{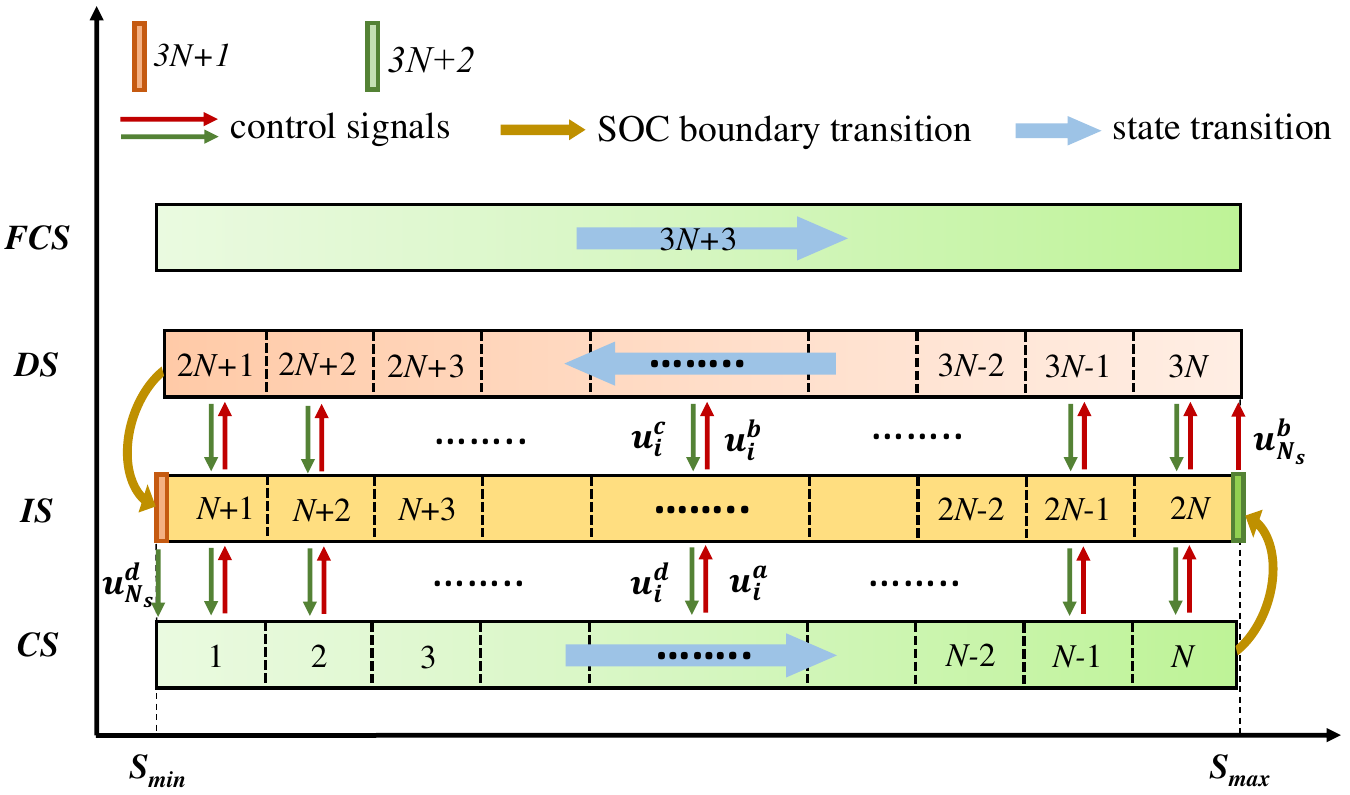}
\vspace{-0.5cm}
\caption{The state transition of aggregated EVs.} 
\label{fig:ControlS}
\end{figure}

\section{Numerical Validation}
In this section, we validate the performance of the eSSM proposed in Section II. The predicted flexibility and power trajectory of aggregated EVs will be compared with the SSM \cite{wang2019state} 
and the baseline individual modeling method (IMM) \cite{zhang2016evaluation}. 
Given the same reference power trajectory, the control performance of different modeling methods will also be tested and compared. \color{black} 
\subsection{Simulation Parameters}
With the population of aggregated EVs to be 
10, 000,  
the characteristic and traveling parameters of aggregated EVs are obtained from \cite{wang2019state}. All EVs' parameters are randomly drawn from the given distributions 
in Table. \ref{tab:characteristic_parameters} and Table. \ref{tab:traveling_parameters}. 
Similar to \cite{wang2019state} 
it is presumed that an EV possesses equivalent rated power for both charging and discharging, along with comparable charging and discharging efficiencies. Upon plugging into the power grid, each EV initiates charging at its rated power without interruption until the maximum SOC is attained or until a control signal prompts a change. 
Most of the EVs start their traveling at about 10:00 am and finish their day of travel to plug in for charging at about 18:00 p.m. 
Our goal is to estimate the flexibility and power trajectory of the aggregated EV based on periodically updated EV parameters.

\begin{table}[t]
\vspace{-0.1cm}
\caption{Characteristic Parameters of Aggregated EVs}
\label{tab:characteristic_parameters}
\centering
\renewcommand{\arraystretch}{1.2}
\setlength\tabcolsep{8pt} 
\begin{tabular}{c c c}
\midrule
\specialrule{0em}{1pt}{1pt}
\midrule
Parameter & Description & Value$^{*}$ \\
\hline
$P_{c}/P_{d}$ & Charging/Discharging Power (kW) & U(5.0, 7.0) \\
$\eta_{c}/\eta_{d}$ & Charging/Discharging Efficiency & U(0.88, 0.95) \\
Q & Battery Capacity (kWh) & U(20.0, 30.0) \\
\midrule
\specialrule{0em}{1pt}{1pt}
\midrule
\end{tabular}
\begin{tablenotes}
\item * $U(a, b)$ denotes a uniform distribution with variation range $[a, b]$.
\end{tablenotes}
\vspace{-0.5cm}
\end{table}

\begin{table}[!htb]
\caption{Traveling Parameters of Aggregated EVs}
\label{tab:traveling_parameters}
\centering
\renewcommand{\arraystretch}{1.2}
\setlength\tabcolsep{8pt}
\begin{tabular}{@{\hspace{0pt}}c@{\hspace{8pt}}c@{\hspace{8pt}}c@{\hspace{0pt}}}
\midrule
\specialrule{0em}{1pt}{1pt}
\midrule
Parameter & Description  & Value$^{*}$ \\
\hline
$S_{s,i}$  & Start Charging SOC  &  $N(0.3, 0.5)\in [0.2,0.4]$ \\
$S_{d,i}$ & Demanded SOC for Travel & $N(0.8, 0.03)\in [0.7, 0.9]$ \\
$t_{s,i}$ & Start Charging Time (h) & $N(-6.5, 3.4)\in [0, 5.5]$ \\
$t_{f,i}$ & Finish Charging Time (h) & $N(8.9, 3.4)\in [0, 20.9]$\\
$S_{min}/S_{max}$ & Minimum/ Maximum SOC Value & 0/1.0\\
\midrule
\specialrule{0em}{1pt}{1pt}
\midrule
\end{tabular}
\begin{tablenotes}
\item * $N(\mu, \delta)$ denotes a normal distribution where $\mu$ is the mean value and $\delta$ is the standard deviation. $[a, b]$ is the variation range.
\end{tablenotes}
\end{table}

\color{black}

For comparison, the individual modeling method (IMM) \cite{zhang2016evaluation} was applied to attain the flexibility and power trajectory of aggregated EVs 
by calculating the sum of all connected EVs in real-time. It has been validated by existing literature \cite{zhang2016evaluation}- \cite{lam2015capacity} that the IMM can provide 
high-accuracy calculation for the flexibility and power trajectory of aggregated EVs. 
Consequently, it is utilized as a baseline to test the accuracy of the proposed eSSM.  

\subsection{Performance of the proposed eSMM Model without Control Commands}
Similar to \cite{wang2020electric}, the number of SOC intervals is set to be 10 (i.e. $N=10$) and the simulation time interval $\Delta t = 15s$. To test the influence of extended states, the SOC states and parameters are updated to the aggregator every $T_{P}=5 min$ period.


With no presence of control signals, every EV gets charged until plugged out or fully charged. 
Fig. \ref{fig:noinput_comparison} shows the power trajectories of aggregated EVs predicted by different modeling methods and the true state distribution over 24 h. The SSM and eSSM both perform well in the prediction of power trajectory for the aggregated EVs. 
Since there are no control commands, EVs are only in CS or $\bm{x}_{3N+2}$ (fully charged).\color{black}
\begin{figure}[!htb]
\centering
\vspace{-0.3cm}
\includegraphics[width=0.95\columnwidth]{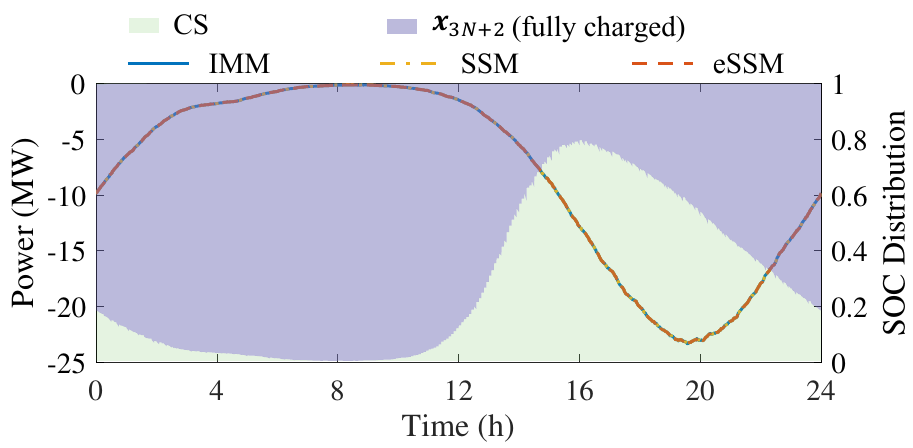}

\caption{Power trajectory prediction of aggregated EVs.}
\label{fig:noinput_comparison}
\end{figure}

The power flexibility calculated by the IMM is shown in Fig. \ref{fig:flexibility_noinput}. (a), serving as the baseline for comparison. The power flexibility estimated by the SSM and the proposed eSSM are presented in  
Fig. \ref{fig:flexibility_noinput}. (b). 
From the figures, it is evident that the proposed eSSM method can give accurate predictions for both the upper bound and the lower bound of the EV flexibility. The SSM method \cite{wang2019state},  nevertheless, fails to give an accurate prediction for the lower bound. 
\begin{figure}[!htb]
    \centering
    \includegraphics[width=1.0\linewidth]{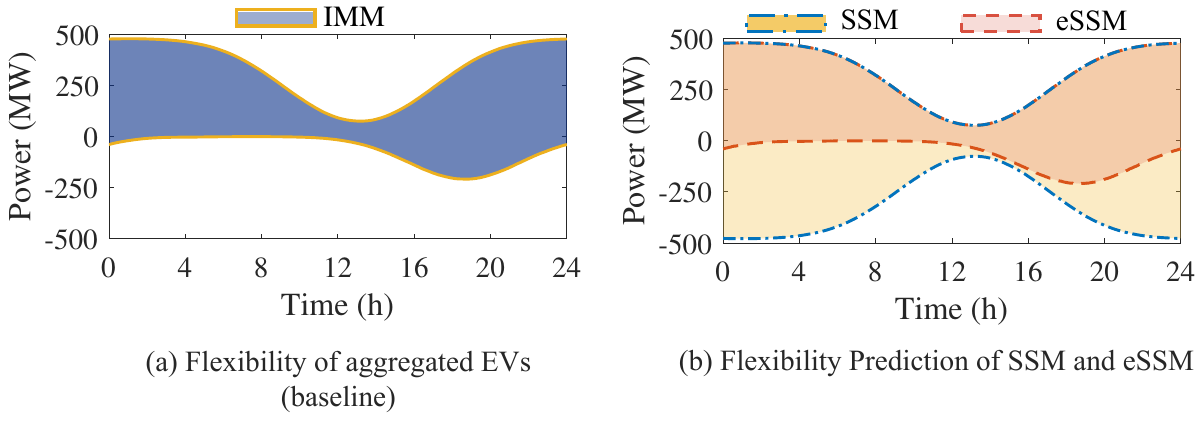}
    \vspace{-0.5cm}
    \caption{A comparison of the flexibility prediction of aggregated EVs by SSM and eSSM, respectively. }
    \label{fig:flexibility_noinput}
    \vspace{-0.3cm}
\end{figure}

This is because when EVs get fully charged but stay connected, they can no longer absorb energy from the grid by switching to CS. In this case, the power trajectory itself represents the lower bound of the EV flexibility. 
However, the SSM incorrectly assumes that 
these EVs can be switched back to CS using a control signal. Since no EVs 
are fully discharged and get into $\bm{x}_{3N+1}$, there is no difference between the upper bounds of SSM and eSSM.

Varying the number of EVs($N_{EV}$), the simulation results under different models are compared in TABLE \ref{tab:power_dif_N}. It indicates that with no control present, the SSM and eSSM hold high accuracy for the prediction of both upper bounds and power trajectories. The prediction accuracy for the lower bounds of the SSM remains low with the variation of EV numbers. In contrast, the eSSM shows consistent 
accuracy for the prediction of both flexibility and power trajectory as $N_{EV}$ varies.
\color{black}
\begin{table}[!htb]
\renewcommand{\arraystretch}{1.2}
\setlength\tabcolsep{8pt}
\caption{Errors of Prediction with Different $N_{EV}$ (\%)}
\vspace{-0.1cm}
\label{tab:power_dif_N}
\centering
\begin{tabular}{c c c c c}
\midrule
\specialrule{0em}{1pt}{1pt}
\midrule
\multirow{2}{*}{$N_{EV}$} & Modeling& \multicolumn{3}{c}
{Estimation Error} \\
\cline{3-5}
& Methods & Upper Bounds&Lower Bounds&Power\\
\hline
\multirow{2}{*}{500} & SSM & 6.78e-15 & 78.61 & 2.84 \\
 & eSSM & 6.78e-15 & 2.84 & 2.84\\
 \hline
  \multirow{2}{*}{5000}& SSM & 3.18e-4 & 71.29 & 2.56 \\
 & eSSM & 3.18e-4 & 2.56 & 2.56\\
 \hline
\multirow{2}{*}{10000}& SSM & 1.11e-3 & 67.66 & 2.87 \\
 & eSSM & 1.11e-3 & 2.87 & 2.87\\
\midrule
\specialrule{0em}{1pt}{1pt}
\midrule
\end{tabular}
\vspace{-0.5cm}
\end{table}

\subsection{Performance of the proposed eSSM with Control Present}
In this section, we will evaluate the 
power control performance using the eSMM and power flexibility estimation when control commands are present. The results will also be compared with the SSM and the baseline IMM. \color{black} Specifically, we consider the frequency regulation control applied in  
\cite{wang2019state}: 
given the target power adjustment, the estimation of EV flexibility, and state distribution, the control center determines the input vectors and control commands, which give the switching probability for EVs in different state intervals. After receiving the control command sent from the control center, the EV will generate a random number $\alpha_{i}$ subjected to the uniform distribution $U(0, 1)$. By comparing $\alpha_{i}$ with the switching probability, the EV determines whether to switch or not. It has been shown in \cite{wang2019state} that this control can achieve real-time power tracking accurately 
for a large population of aggregated EVs. Interested readers can refer to \cite{wang2019state} for more details. \color{black}

Given a reference power trajectory with random large disturbances occurring every 3 hours (see Fig. \ref{fig:power_ref}), 
Fig. \ref{fig:input_comparison} shows the control results of the three methods and the variation in state distribution over time. It demonstrates that the eSSM control method responds accurately to the power adjustment requirements. 
Nevertheless, the SSM may fail to respond accurately to the power requirements. For example, Fig. \ref{fig:control_detail} details the control results 
of the aggregated EVs at 6:00 am and 18:00 pm.
\color{black}For the power consumption command at 6:00 am, the control using the SSM fails to respond to the requirement, because it tries to control the EVs in $\bm{x}_{3N+2}$ (fully charged) back to CS while in reality, they cannot absorb energy from the grid anymore. For the power provision command at 18:00 pm, it incorrectly assumes that the EVs in $\bm{x}_{3N+1}$ (fully discharged) can be switched to DS.
\begin{figure}[!htb]
    \vspace{-0.3cm}
    \centering
    \includegraphics[width=0.95\linewidth]{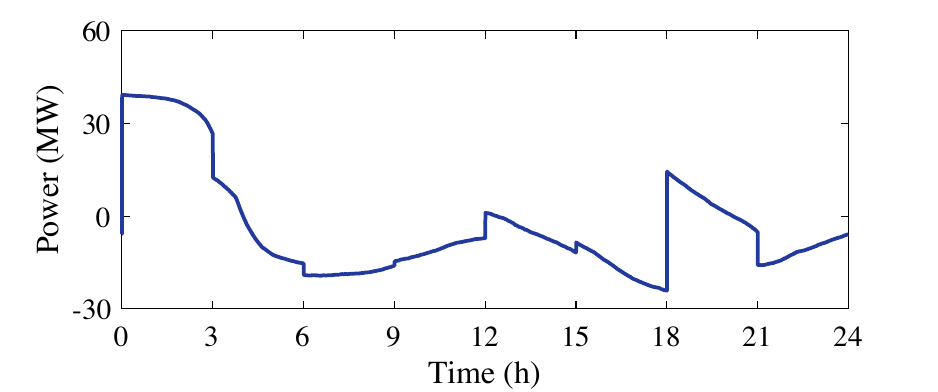}
    \caption{Reference power trajectory}
    \label{fig:power_ref}
\end{figure}
\begin{figure}[!htb]
\centering
\includegraphics[width=0.9\columnwidth]{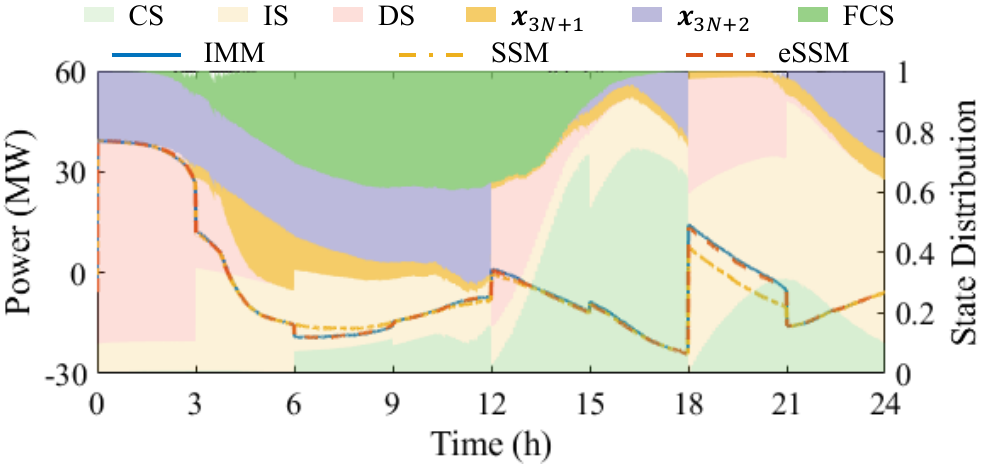}
\caption{Power profiles of EVs responding to power requirements.}
\label{fig:input_comparison}
\end{figure}
\begin{figure}[!htb]
\vspace{-0.2cm}
    \centering
    \includegraphics[width=0.95\linewidth]{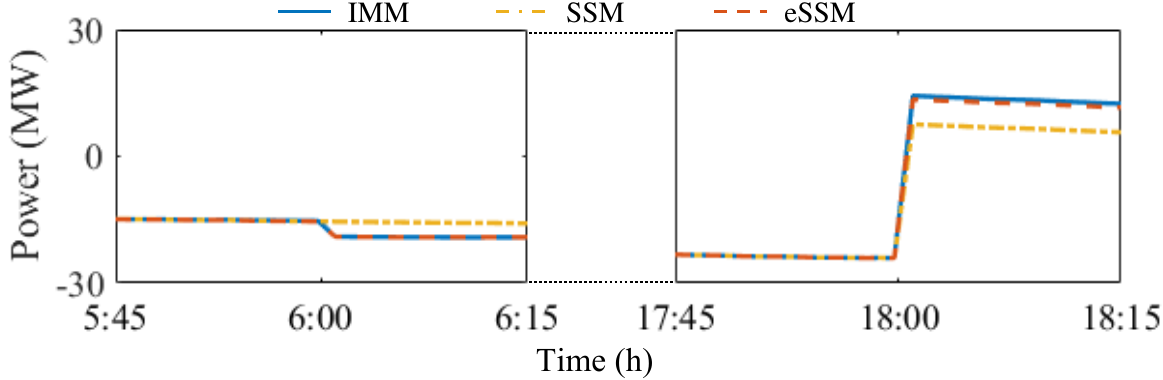}
    \caption{Response to power requirements of different modeling methods.}
    \label{fig:control_detail}
    \vspace{-0.5cm}
\end{figure}

Fig. \ref{fig:flexibility_input} depicts the flexibility of the aggregated EVs 
estimated by the three modeling methods. From 3:00 to 12:00, there is a significant difference between the upper bounds of SSM and IMM. This is because, after the power provision 
command at 00:00, a large amount of EVs switch from IS to DS. They continued to be discharged and were fully discharged ($\bm{x}_{3N+1}$) at about 3:00 (see Fig. \ref{fig:input_comparison}). 
Similar poor accuracy for the lower bound estimated by the SSM can be observed as well. It is because there is still a significant amount of fully charged  EVs ($\bm{x}_{3N+2}$) between 0:00 to 12:00 as seen from Fig. \ref{fig:input_comparison}, and the SSM incorrectly assumes that fully charged EV can be switched to CS. 
Situations are similar for 21:00-24:00. From these analyses, we can see that 
the prediction error of the lower bound by the SSM increases as the proportion of fully charged EVs $\bm{x}_{3N+2}$ increases, whereas 
the prediction error of the upper bound by the SSM increases as the proportion of fully discharged EVs ($\bm{x}_{3N+1}$) increases. 
In contrast, the prediction results of the eSSM align well with the baseline IMM, showing high accuracy in estimating the flexibility of aggregated EVs when control commands are present.  

\begin{figure}[!htb]
    \centering
    \includegraphics[width=1.0\linewidth]{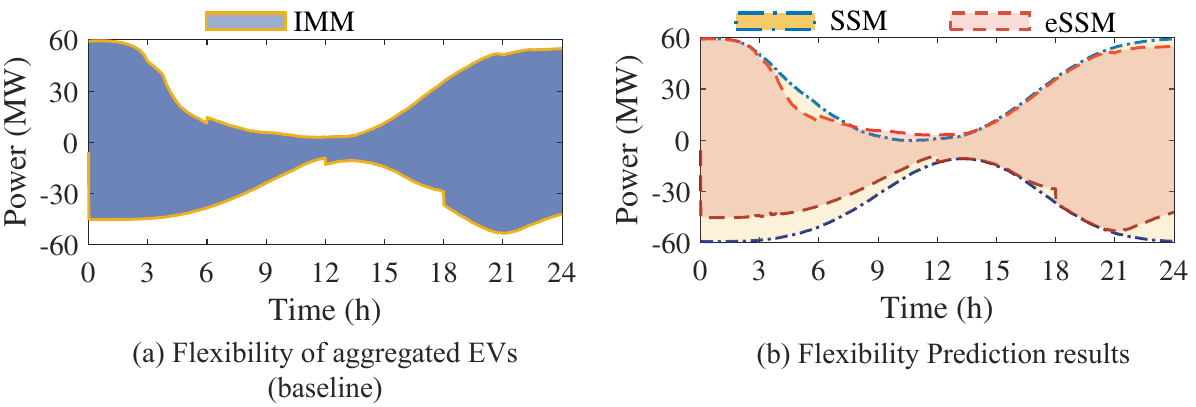}
    \caption{A comparison of the flexibility prediction of aggregated EVs when control present.}
    \label{fig:flexibility_input}
\end{figure}
\vspace{-0.3cm}

\color{black}

\section{Conclusion}
In this paper, we proposed an extended state space model (eSSM) to predict the flexibility and power trajectories of aggregated EVs. Control methods based on the eSSM are developed for effective power reference tracking to provide ancillary service. The conclusions are summarized as follows:

(1) Compared with traditional SSM, the proposed eSSM provides more accurate predictions for the flexibility and power trajectory of aggregated EVs. This improvement is achieved by adding two boundary intervals to incorporate the limited flexibility of fully charged and discharged EVs.

(2) When responding to power adjustment requirements of ancillary service, the power control based on the eSSM demonstrates high control accuracy.

\color{black}



\bibliographystyle{IEEEtran}
\bibliography{main}

\end{document}